\documentclass{PoS}
\usepackage{epsfig}
\usepackage{cite}

\title{Topological properties of the SU(3) random vortex world-surface model}

\ShortTitle{Topological properties of the SU(3) random vortex world-surface
model}

\author{\speaker{Michael Engelhardt}%
         \thanks{Supported by the U.S.~DOE under grant DE-FG02-96ER40965.}\\
        Department of Physics, New Mexico State University, Las Cruces,
        NM 88003, USA\\
        E-mail: \email{engel@nmsu.edu}}

\abstract{The random vortex world-surface model is an infrared effective
model of Yang-Mills dynamics based on center vortex degrees of freedom.
These degrees of freedom carry topological charge through writhe and
self-intersection of their world-surfaces. A practical implementation
of the model realizes the vortex world-surfaces by composing them of
elementary squares on a hypercubic lattice. The topological charge for
specifically such configurations is constructed in the case of SU(3)
color. This necessitates a proper treatment of vortex color structure
at vortex branchings, a feature which is absent in the SU(2) color case
investigated previously. On the basis of the construction, the topological
susceptibility is evaluated in the random vortex world-surface ensemble,
both in the confined low-temperature as well as in the deconfined
high-temperature phase.}

\FullConference{8th Conference Quark Confinement and the Hadron Spectrum\\
                 September 1-6, 2008\\
                 Mainz. Germany}

\begin{document}

\section{Introduction}
The random vortex world-surface model 
\cite{m1,m2,m3,su3conf,su3bary,su3freee,su4,sp2} is an infrared effective
description of the strong interaction vacuum which starts from the
assumption that the relevant gluonic degrees of freedom are center vortices.
These are tubes of chromomagnetic flux which are envisioned to be percolating
throughout space. The flux tubes are closed, since chromomagnetic flux must
be continuous in accordance with the Bianchi constraint, and their flux is
quantized in a way which is determined by the center of the gauge group.
Adopting a four-dimensional (Euclidean) space-time point of view, an
ensemble of random vortex world-surfaces is constructed (in practice,
employing Monte-Carlo methods), and relevant observables are then
calculated within this ensemble.

The physical picture of the strong interaction vacuum sketched above,
motivating the random vortex world-surface model, was originally proposed
as an explanation of the confinement phenomenon \cite{hooft,aharonov}.
The percolating random distribution of magnetic flux sufficiently
disorders Wilson loops such as to generate an area law for them.
In the course of subsequent investigations
\cite{jg3,jg2,df1,df2,vorint,per,rb,hoellw} sparked by the
development of techniques which permit the detection and study of
center vortices in lattice gauge configurations \cite{jg2,df2}, a more
comprehensive picture emerged. Not only does the vortex vacuum
generate confinement, but it also can account for the other two
core characteristics of the strong interaction, namely, the
spontaneous breaking of chiral symmetry, and the axial $U_A (1)$
anomaly.

The present investigation further contributes to this broadening of the
scope of the vortex picture. While the confinement properties of the
random vortex world-surface model have been investigated in a variety
of settings \cite{m1,su3conf,su3bary,su3freee,su4,sp2}, the topological
and chiral properties had hitherto only been studied for the simplest
case of $SU(2)$ color \cite{m2,m3}. Here, the topological properties are
investigated for $SU(3)$ color. After dealing with some new subtleties
concerning vortex color structure which are not present in the $SU(2)$
case, the topological susceptibility of the $SU(3)$ random vortex
world-surface ensemble is evaluated as a function of temperature,
including both the confined as well as the deconfined phase.

\section{Color structure of SU(3) center vortices and topological charge}
\label{colorsec}
The chromomagnetic flux carried by a center vortex is measured by Wilson
loops $W_C $ encircling the vortex along paths $C$. Vortex flux quantization
expresses itself in the fact that such Wilson loops take only (nontrivial)
values in the center of the gauge group, for $SU(3)$,
\begin{equation}
W_C = \frac{1}{3} \, \mbox{Tr} \, {\cal P} \exp \left(
i\oint_{C} A_{\mu } dx_{\mu } \right) = \exp (\pm 2\pi i/3) \ .
\label{fluxquant}
\end{equation}
Since the two center elements in (\ref{fluxquant}) are related by complex
conjugation, the corresponding vortex fluxes are related by inversion of
their space-time orientation. There is, therefore, only one type of
vortex in the $SU(3)$ case. Nevertheless, vortex flux can branch; the
Bianchi constraint admits the possibility of a flux associated with
the center phase $\exp (-2\pi i/3) = \exp (4\pi i/3)$ splitting into
two oppositely oriented fluxes, each associated with the center phase
$\exp (2\pi i/3)$.

Evaluating the topological charge
\begin{equation}
Q=\frac{1}{32\pi^{2} } \int d^4 x \, \epsilon_{\mu \nu \lambda \tau } \
\mbox{Tr} \ F_{\mu \nu } F_{\lambda \tau }
\label{topcharge}
\end{equation}
calls for a representation of vortex flux via the algebra of the gauge
group. This is conveniently achieved by adopting an Abelian gauge, in
which the gauge field $A_{\mu } $ and its field strength $F_{\mu \nu } $
are proportional to diagonal matrices $T$ in color space. Contrary to
the $SU(2)$ case, in which there is no residual (gauge) freedom in the
set of matrices $T$ entering the description, for $SU(3)$, there is
such a choice. In order to generate both nontrivial center elements
when evaluating Wilson loops,
\begin{equation}
W=(1/3) \, \mbox{Tr} \, \exp (2\pi iT/3) = \exp (\pm 2\pi i/3) \ ,
\end{equation}
one minimally needs a set with two elements, e.g.,
$T \in \{ \pm \, \mbox{diag} \, (1,1,-2) \} $.
However, such a choice has the drawback that vortex branchings necessarily
carry Abelian magnetic monopoles (why this is a drawback will be explained
further below): If a vortex associated with the center element
$\exp (-2\pi i/3)$ splits into two oppositely oriented vortices, flux
is only continuous modulo $2\pi $ in each color component,
\begin{equation}
\frac{2\pi }{3} \, \mbox{diag} \, (-1,-1,2) \ = \
2\cdot \frac{2\pi }{3} \, \mbox{diag} \, (1,1,-2) \ - \
2\pi \, \mbox{diag} \, (1,1,-2) \ .
\end{equation}
In order to always be able to deform Abelian magnetic monopole
world-lines\footnote{Note that magnetic monopole world-lines generally
cannot be eliminated altogether due to the nonorientability of generic
world-surfaces.} away from vortex world-surface branchings, a more
flexible basis is necessary \cite{cw1}, such as
\begin{equation}
T \in \left\{ \pm \, \mbox{diag} \, (1,1,-2) \, , \
\pm \, \mbox{diag} \, (1,-2,1) \, , \ \pm \, \mbox{diag} \, (-2,1,1) \,
\right\} \ .
\label{nonminchoice}
\end{equation}
A description of vortex surfaces in terms of patches which are each
associated with one of the elements in (\ref{nonminchoice}) is well-suited
for the evaluation of topological charge.

According to (\ref{topcharge}), topological charge density is generated
when field strength components $F_{\mu \nu } $, $F_{\lambda \tau } $
exist at a given space-time position such that $\mu , \nu , \lambda , \tau $
span all four space-time dimensions. Since a vortex world-surface is
associated with a field strength component $F_{\mu \nu } $ such that
$\mu , \nu $ denote the directions perpendicular to the
world-surface\footnote{E.g., a static chromomagnetic flux in 3-direction,
which corresponds to a world-surface extending in the 3- and 4-directions,
carries a 3-component of the magnetic field, $B_3 $, and thus a field
strength component $F_{12} $.}, this translates into the statement that
the vortex world-surfaces present at a given space-time position must
span all four space-time dimensions. There are two ways in which this
can take place, namely, if two distinct surfaces intersect, or if one
surface writhes such that it explores all four directions
\cite{m2,cw1,contvort,bruck,jordan}.

\begin{figure}
\vspace{-0.3cm}
\centerline{
\epsfig{file=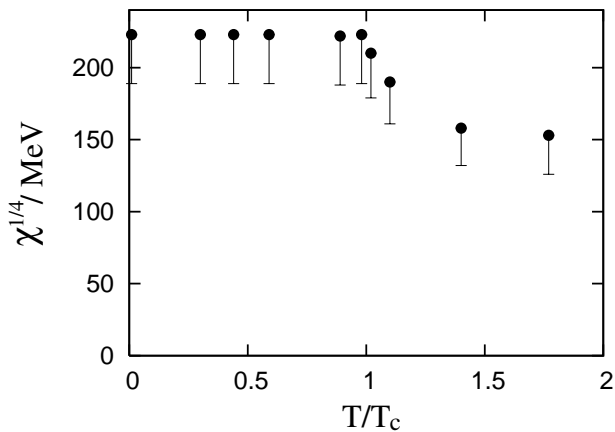,width=6cm,angle=-90}
}
\caption{Fourth root of the topological susceptibility measured in the
$SU(3)$ random vortex world-surface model, as a function of temperature.
In the employed random surface ensemble, residual statistical uncertainties
are smaller than the filled circle symbols depicting the measured data;
the origin of the displayed systematic downward uncertainty is explained
in the main text.}
\label{chifig}
\end{figure}

To have a practicable scheme of generating an ensemble of
random vortex world-surfaces, the latter are composed of elementary
squares on a hypercubic lattice. In such a setting, topological charge
density only occurs at lattice sites. This simplified
description comes, however, at a price. Due to the discrete set of
directions in which surfaces can extend, ambiguous features appear in
the configurations, features which would merely constitute a
negligible set of measure zero if one considered arbitrary surfaces
in continuous space-time. For one, two surfaces may coincide along whole
line segments in space-time instead of only discrete intersection points.
In this case, there is generally no unambiguous assignment of the
participating vortex elementary squares to two distinct surfaces. To
remedy this, vortex surfaces are transferred to a finer
lattice and slightly deformed (by less than half the original lattice
spacing) until all ambiguities are removed. While this permits the
assignment of a topological charge, the described deformation procedure
can introduce
\pagebreak
additional spurious topological charge density at
the fine lattice scale, leading to an overestimate of the topological
susceptibility. This in fact constitutes the main source of error
in the measurement. The corresponding downward uncertainty in the data
shown in Fig.~\ref{chifig} is estimated by rescaling with the
(appropriate power of the) vortex density, which also is increased by
the deformation procedure in a way which is expected to roughly
track the spurious increase in the topological charge density.

A further ambiguity present in the configurations is related to
Abelian magnetic monopole world-lines. If tied rigidly to
vortex branchings, they necessarily run through lattice sites, precisely
where topological charge density is concentrated. Such singular coincidences
also preclude an unambiguous assignment of topological charge. They are
removed by deforming monopole world-lines away from lattice sites,
taking advantage of the flexible color structure defined by
(\ref{nonminchoice}).

\section{Random vortex world-surface model}
As already mentioned above, and described in detail in \cite{m1,su3conf},
a practical Monte-Carlo scheme for generating a random vortex world-surface
ensemble is arrived at by composing the surfaces from elementary squares
on a hypercubic lattice. The lattice spacing is a fixed physical quantity
related to the vortex thickness, which defines the ultraviolet cutoff of
the model; it amounts to $0.39\, \mbox{fm} $ if one sets the scale via
the zero-temperature string tension, $\sigma = (440\, \mbox{MeV} )^{2} $.
The ensemble is weighted by an action penalizing vortex world-surface
curvature,
\vspace{-0.4cm}

\begin{figure}[h]
\centerline{\hspace{0.1cm}
\epsfig{file=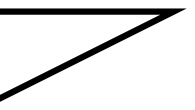,width=1.2cm} }
\end{figure}
\vspace{-1.8cm}

\begin{equation}
\hspace{-3cm} S\ \ = \ \ c \ \times
\end{equation}
\vspace{-0.4cm}

\noindent
i.e., each adjacent, non-coplanar pair of vortex squares adds an increment
$c$ to the action. The coefficient $c$ is tuned to reproduce the
value $T_c /\sqrt{\sigma } =0.63$ obtained in $SU(3)$ lattice Yang-Mills
theory, where $T_c $ denotes the deconfinement temperature\footnote{The
temperature in the model is varied as usual, via the Euclidean time extent
of the lattice.}. This yields $c=0.21$. Evaluating the topological charge
$Q$ of configurations as indicated in the previous section yields the
result shown in Fig.~\ref{chifig} for the topological susceptibility
$\chi = \langle Q^2 \rangle /V$ (where $V$ denotes the lattice four-volume).

\section{Conclusions}
The raw data depicted in Fig.~\ref{chifig} lie somewhat above analogous
results obtained in $SU(3)$ lattice Yang-Mills theory. The latter are
concentrated around $\chi^{1/4} \approx 190\, \mbox{MeV} $, although
there is a considerable spread in reported values, cf.~in particular
Table 1 in \cite{panag}. For finite temperatures, cf., e.g., \cite{alles}.
However, as indicated in section \ref{colorsec}, ambiguities present
in the vortex configurations due to the specific hypercubic realization
of their world-surfaces employed in this work leave a substantial
systematic downward uncertainty in the results, as shown in
Fig.~\ref{chifig}. Taking this uncertainty into consideration,
it cannot be excluded that the $SU(3)$ random vortex world-surface
ensemble possesses a topological susceptibility compatible with the
$SU(3)$ Yang-Mills one. To resolve this uncertainty, a more
flexible construction of the world-surfaces, allowing them to
extend into arbitrary directions in space-time, is needed, e.g., a
construction in terms of random triangulations.

\end{document}